\documentclass[12pt]{article}
\usepackage{amssymb}
\usepackage{graphicx}
\usepackage[normal]{caption2}
\begin{document}
\section*{On an alternative explanation of anomalous scaling and how well-defined is the concept of inertial range}

\bigskip M. Kholmyansky$^{1 }$, A. Tsinober$^{1 ,2,}$
\footnote{ Author to whom correspondence should be addressed;
electronic mail: a.tsinober@imperial.ac.uk;
tsinober@eng.tau.ac.il}

$^{1}$ {\it Department of Fluid Mechanics and Heat Transfer,
Faculty of Engineering, Tel-Aviv University, Tel-Aviv 69978,
Israel}

$^{2}$ {\it Institute for Mathematical Sciences and Department of
Aeronautics, Imperial College, SW7 2PG London, United Kingdom}
\vskip 0.5cm
\begin{center}
\begin{minipage}{0.9\textwidth}
{\rm Abstract.} {\small The main point of this communication is that there is a small 
non-negligible amount of eddies-outliers/very strong events (comprising a significant subset 
of the tails of the PDF of velocity increments in the nominally-defined inertial range) for which 
viscosity/dissipation is of utmost importance at whatever high Reynolds number. These events contribute 
significantly to the values of higher-order structure functions and their anomalous scaling. Thus the anomalous scaling is not an attribute of the conventionally-defined inertial range, and the latter is not a well-defined concept. The claim above is supported by an analysis of high-Reynolds-number flows in which among other things it was possible to evaluate the instantaneous rate of energy dissipation.}
\end{minipage}
\end{center}
\bigskip

\emph{There are a variety of models of higher statistics that have meager or nonexistent deductive support from the NS equations but can be made to give good fits to experimental measurements}$^{\ref{ref:gokr}}$. These include `explanations' of what is called anomalous scaling observed experimentally
for higher-order structure functions of velocity and temperature increments,
such that their scaling exponents $\zeta_{p}=p/3-\mu_{p}<p/3$ are
nonlinear concave functions of the order $p$. Starting with refined
similarity hypotheses by Kolmogorov$^{\ref{ref:kolm}}$ and Oboukhov$^{\ref{ref:obou}}$, numerous
phenomenological models have been proposed to describe these deviations
considered as the major manifestation of intermittency in the inertial range$^{\ref{ref:frish}-\ref{ref:ssya}}$. The dominant of these models has been the multi-fractal formalism$^{\ref{ref:frish}}$, others claimed the Reynolds number dependence as responsible$^{\ref{ref:lund}-\ref{ref:qian}}$.
The common in all these approaches is the basic, widely accepted premise
that \emph{in the inertial range, the viscosity plays in principle no role}$^{\ref{ref:ruel}}$ so that \emph{nonlinear dependence of the algebraic scaling exponents} $\zeta_{p}$ \emph{on the moment order p is a manifestation of
the inertial-range intermittency}$^{\ref{ref:ssya}}$ with the \textit{inertial range}
defined as $\eta \ll r\ll L$ (with $\eta $ being the Kolmogorov and $L$ --- some
integral scale). Thus the issue is directly related to what is called
inertial (sub)range and how inertial it is.

The main point of this communication is that there is a small non-negligible
amount of eddies-outliers/very strong events (comprising a significant
subset of the tails of the PDF of $\Delta u_{i}(r)$ in the \textit{nominally} defined inertial range 
$\eta \ll r\ll L$) for which viscosity/dissipation
is of utmost importance at whatever high Reynolds number. In other words,
the inertial range is ill-defined in the sense that not all, but almost all
statistics of $\Delta u_{i}(r)$ is independent of viscosity. As long as
one deals with low-order statistics of $\Delta u_{i}(r)$ (as Kolmogorov did)
this is of little (but not always negligible) importance. However, it
appears that these events contribute significantly to the higher-order
structure functions and thereby a non-negligible contribution to the higher-order structure functions is dominated by viscosity. In other words, the
`anomalous scaling' as exhibited by the behavior of higher-order structure
functions is to a large extent due to significant contribution of
viscosity/dissipation in the \emph{inertial range} as commonly defined. The
higher the order of the structure function, the stronger is the contribution
due to viscosity (i.e. from the tails of the PDFs of $\Delta u_{i}(r)$) and
the weaker is the `inertial' contribution (i.e. from the core of those PDFs) to the structure function. Thus it seems problematic to speak about inertial-range behavior of higher-order structure functions.

The support for the above view comes from a recent analysis of high-Reynolds-number data in field experiments$^{\ref{ref:gkk},\ref{ref:kty}}$. The experimental facilities and related matters are described in these papers and references therein. We give here a very brief reminding on these.

The measurement system, developed by the group of Prof. Tsinober,
consists of the multi-hot-wire probe connected to the anemometer
channels, signal normalization device (sample-and-hold modules and
anti-aliasing filters), data acquisition and calibration unit. The
probe is built of five similar arrays. Each calibrated array
allows to obtain three velocity components ``at a point''. The
differences between the properly chosen arrays give the tensor of
the spacial velocity derivatives (without invoking of Taylor
hypothesis), temporal derivatives can be obtained from the
differences between the sequential samples.

\begin{figure}[t]
\begin{center}
\includegraphics[bb=0 0 540 340,width=.45\textwidth]{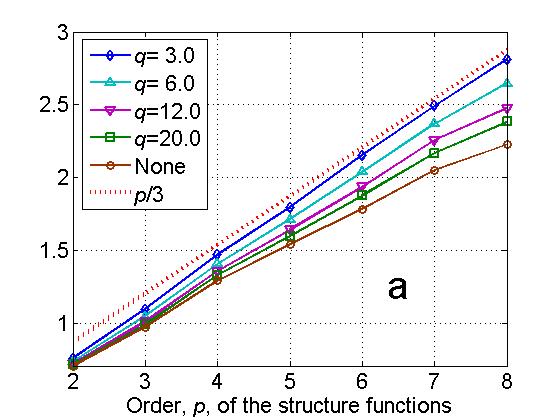}
\includegraphics[bb=0 0 540 340,width=.45\textwidth]{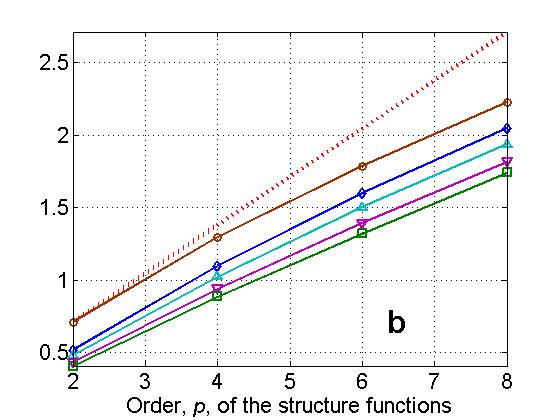}
\caption{{\protect\small {(a) Scaling exponents of structure
functions at $Re_{\lambda}\sim 10^{4}$ for the longitudinal velocity component corresponding to the full data and the same data in which the strong dissipative events (when at least at one
point $x$ or $x+r$ the instantaneous dissipation $\epsilon > q\langle\epsilon\rangle$) with various
thresholds $q$ were removed. (b) Scaling exponents for the strong events themselves.}}} 
\label{fig:scexp}
\end{center}
\end{figure}

\begin{figure}[t]
\begin{center}
\includegraphics[bb=0 0 840 340,width=\textwidth]{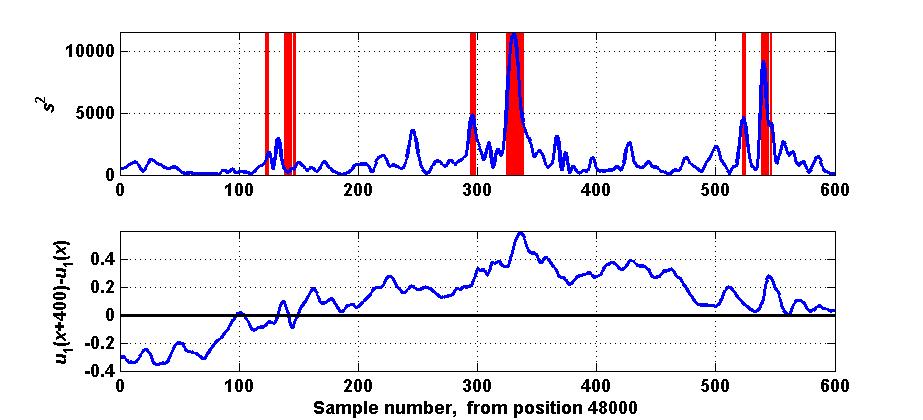}
\caption{{\protect\small {Example of simultaneous time series of the squared magnitude of the rate of strain tensor, $s^2$, proportional to the dissipation $\epsilon$ (top) and the velocity increments, 
$\Delta u_{1}\equiv u_{1}(x+r)-u_{1}(x)$ for $r=400\eta$ (bottom). The marked segments correspond to the strong events, selected with the value of the threshold $q=12$ equivalent to the value of $s^2\approx 4,000$. It should be noted that the first two marked segments are considered strong events because the value of $s^2$ reaches the threshold at the point $x+r$.}}} 
\label{fig:timeser}
\end{center}
\end{figure}
\newpage

\begin{figure}[t!]
\begin{center}
\includegraphics[bb=0 0 540 400,width=.45\textwidth]{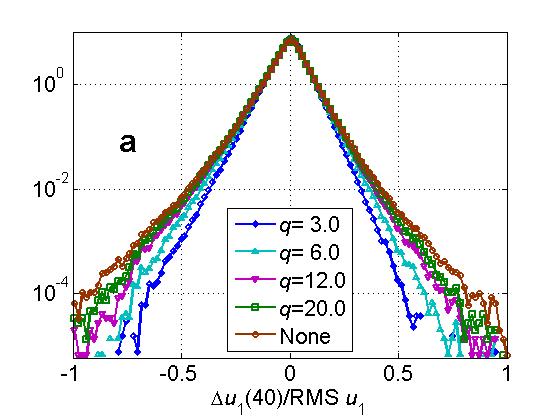}
\includegraphics[bb=0 0 540 400,width=.45\textwidth]{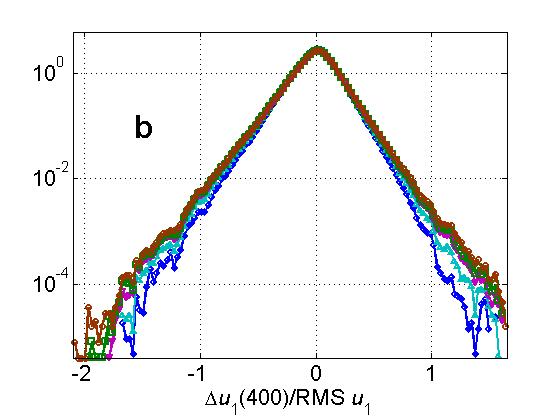}
\caption{{\protect\small {PDFs of the increments of the longitudinal velocity
component for the same data as in Fig. \ref{fig:scexp}. (a) $r/\protect\eta =40$ corresponds to the
lower edge of the inertial range. (b) $r/\protect\eta =400$ is deep in the inertial range.}}} 
\label{fig:pdfs}
\end{center}
\end{figure}

\begin{figure}[h!]
\begin{center}
\includegraphics[bb=0 0 540 400,width=.45\textwidth]{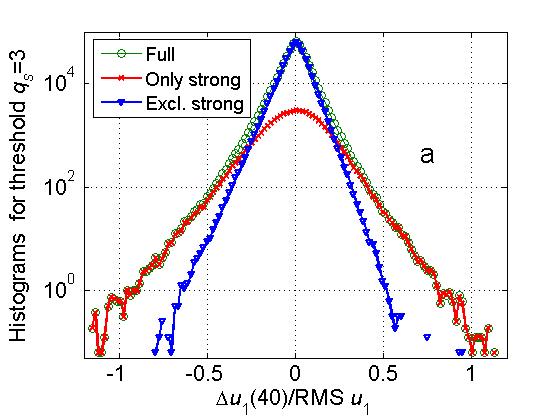}
\includegraphics[bb=0 0 540 400,width=.45\textwidth]{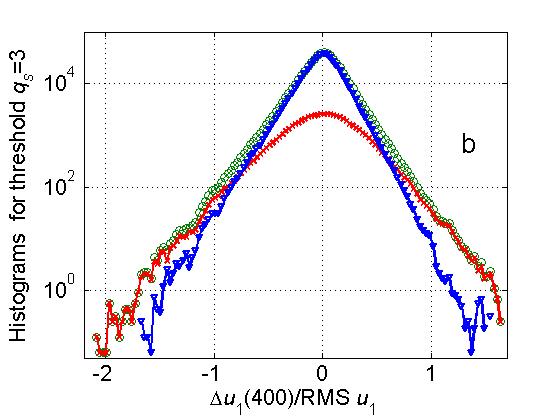}
\caption{{\protect\small {Histograms of the increments of the longitudinal velocity
component for the same data as in Fig. \ref{fig:scexp} for the threshold $q=3$. $r/\protect\eta =40$ (a). $r/\protect\eta =400$ (b).}}} 
\label{fig:hists}
\end{center}
\end{figure}
The calibration unit produces a jet with variable velocity
magnitude and variable angles around two orthogonal axes. The
probe is located in the jet core, the values of velocity magnitude
and angles are recorded together with the readings of the
anemometer channels and later approximated by polynomials, used
for obtaining velocity components from the measured hot-wires
data.

\begin{figure}[t]
\begin{center}
\includegraphics[bb=0 0 640 440,width=.45\textwidth]{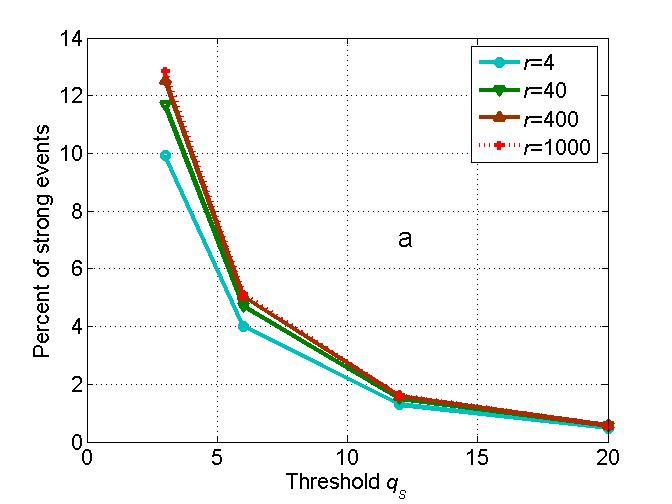}
\includegraphics[bb=0 0 640 440,width=.45\textwidth]{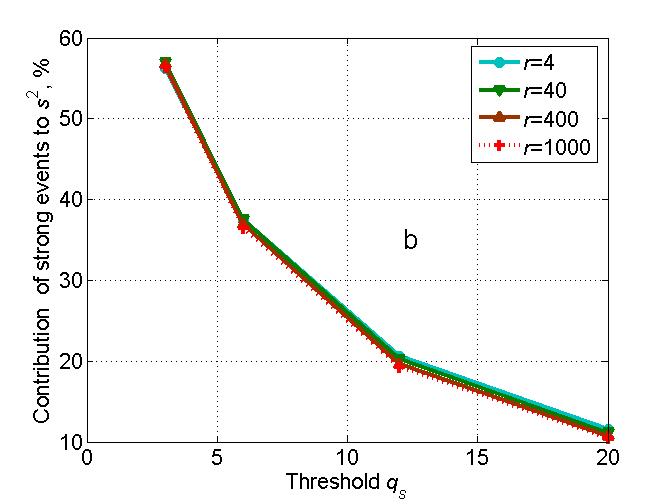}
\caption{{\protect\small {Percent of the strong dissipative events as defined in the
text (a) and their contribution to the total dissipation (b) as a
function of the threshold $q$ for various separations $r$.}}} 
\label{fig:strain}
\end{center}
\end{figure}

\begin{figure}[t]
\begin{center}
\includegraphics[bb=0 0 540 440,width=.3\textwidth]{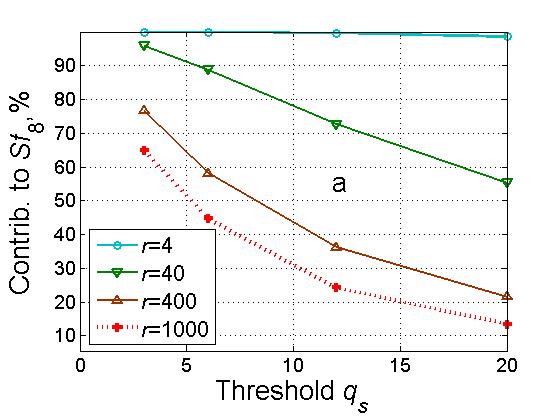}
\includegraphics[bb=0 0 540 440,width=.3\textwidth]{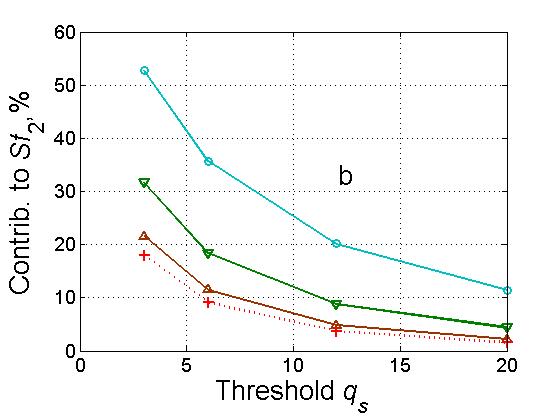}
\includegraphics[bb=0 0 540 440,width=.3\textwidth]{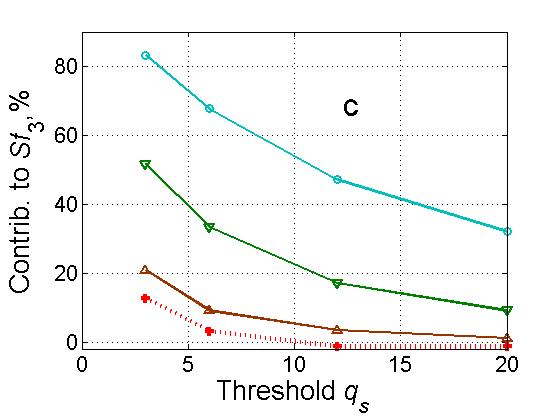}
\caption{{\protect\small {Contributions of the strong dissipative events, as defined in the
text, to the eighth-order structure function (a), the second-order
structure function (b) and the third-order structure function (c) as a
function of the threshold $q$ for various separations $r$.}}} 
\label{fig:du}
\end{center}
\end{figure}

A selection of results is shown in Fig. \ref{fig:scexp} --- Fig. \ref{fig:du}.

Fig. \ref{fig:scexp} (a) shows the scaling exponents of structure
functions $S_{p}^{||}(r)$ up to order 8 corresponding to the full data and
the same data in which the strong dissipative events with various
thresholds were removed. By an event we mean here a velocity increment, 
$\Delta u_{1}\equiv u_{1}(x+r)-u_{1}(x)$. It is qualified as a strong
dissipative event if at least at one of its ends $(x,x+r)$ the instantaneous
dissipation $\epsilon > q \langle \epsilon \rangle$ for $q>1$. We have chosen 
$q=3,6,12$ and $20$. This corresponds to the instantaneous Kolmogorov-like
scales $0.76$, $0.64$, $0.54$ and $0.47$ of the conventional Kolmogorov
scale $\eta$ based on the mean dissipation $\langle \epsilon \rangle$. It
is seen that the removal of the strong dissipative events results in an increase
of the exponents $\zeta_{p}$. For example, with the removal of the dissipative
events between the threshold $3\langle \epsilon \rangle$ ($0.76\eta$) and 
$6\langle \epsilon \rangle$ ($0.64\eta$) the dependence of $\zeta_{p}$ on 
$p$ becomes pretty close to the Kolmogorov $p/3$. The strong events/outliers
themselves have different scaling properties (Fig. \ref{fig:scexp} (b)). The time series in Fig. \ref{fig:timeser} illustrate the selection of strong events.

The next example in Fig. \ref{fig:pdfs} shows that indeed the removal of the strong dissipative
events results in narrowing of the tails in the PDFs of $\Delta u_{1}(r)$.

As an additional illustration we show in Fig. \ref{fig:hists} two examples of histograms of the increments of $u_1$ for the whole field, with removed strong dissipative events for the threshold $q=3$ and the dissipative events themselves for the same threshold, for the same data as in Fig. \ref{fig:scexp}.

The effect of the removal of the strong dissipative
events is obviously much stronger for higher-order structure functions. 
For example, there are only $5\%$ of dissipative events (Fig. \ref{fig:strain} (a)) 
for $q=6$ sitting mostly at tails
of the PDF of $\Delta u_{i}(r)$ for $r/\eta =400$ (i.e. deep in the
`inertial' range), which contribute about $36\%$ to the total dissipation
(Fig. \ref{fig:strain} (b)). These events contribute nearly $60\%$ to the value of
$S_{8}^{\shortparallel}(r)$ at $Re_{\lambda}\sim 10^{4}$ (Fig. \ref{fig:du} (a)). These same
events change the $S_{2}^{\shortparallel}(r)$ by about $11.5\%$ (Fig. \ref{fig:du} (b)), but
contribute about $9\%$ to $S_{3}^{\shortparallel}(r)$ (see Fig. \ref{fig:du} (c)).

It is noteworthy that the data used here$^{\ref{ref:gkk},\ref{ref:kty}}$ was somewhat spatially
underresolved, $1\div 3\eta$. This means that the conclusions are to some
extent qualitative. However, with properly resolved data the strong
dissipative events, lost in the underresolved ones, would somewhat enhance the
tendencies just described above. This is in agreement with the fact that
essentially the same results are obtained using the same data smoothed over
up to eight sequential samples. Additional support comes from reference$^{\ref{ref:ssya}}$, 
indicating that the underresolved data reproduce
faithfully the flow at scales about two times smaller than those resolved 
($\sim 0.6\eta$) at least as concerns the instantaneous dissipation rate.
Finally, using enstrophy $\omega^{2}$ and/or the surrogate $(\partial
u_{1}/\partial x_{1})^{2}$ as a criterion for the threshold instead of the
true dissipation $\epsilon$ gives the same qualitative (but not
quantitative) results.

Summarizing, the main point is the distinction between roughly two kinds of
events (in the \textit{nominal} inertial range $\eta \ll r\ll L$),
contributing to the value of the velocity increment. One is represented by
the core of the PDF of $\Delta u_{i}$, and the other --- by the outliers/extreme
events (comprising a significant subset of the tails of the PDF of $\Delta
u_{i}$). They have not only different statistical properties (such as scaling if
such exists), but also are of different nature in the sense that the former
exhibit `inertial' behavior as reflected in the slopes of low-order
structure functions, whereas the latter are dominated by viscous effects as
seen in the slopes of higher-order structure functions. Removal of these
highly-dissipative events brings the dependence of $\zeta_{p}$ on $p$ 
pretty close to the Kolmogorov $p/3$. Thus the anomalous scaling is not
the attribute of the inertial range. Our results leave little doubt that the
strong dissipative events contribute significantly to the anomalous scaling
of higher-order structure functions. However, there are other effects which
are expected to contribute to `anomalous scaling' such as a variety of
nonlocal effects understood in a broad sense as direct and bidirectional
coupling/interaction between large and small scales$^{\ref{ref:atsi},\ref{ref:khts}}$. 
The quality of our data does not allow to address properly this and similar issues. This is
a matter of far more precise and well-controlled experiments which among
other things require information at high Reynolds numbers with sub-Kolmogorov resolution.

Along with the fact that velocity increments (let alone structure functions
and their scaling if such exists) are not the only objects of interest and 
\emph{do not constitute a representation basis for a flow}$^{\ref{ref:gokr}}$,
they are not a good object to define a perfect inertial range. Such a definition seems
to be not possible in principle due to a variety of nonlocal effects as
mentioned above$^{\ref{ref:atsi},\ref{ref:khts}}$.

A special remark is about the contribution of the dissipative events as
defined/described above to the 4/5 law. These events \textit{do} contribute
to the 4/5 law, and removing them leads to an increase of the scaling
exponent above unity, see Fig. \ref{fig:scexp} (a) and Fig. \ref{fig:du} (c). An important
point here is that the neglected viscous term in the Karman--Howarth equation
does not contain all the viscous contributions. Those which are present in
the structure function $S_{3}$ itself remain and keep the 4/5 law precise.
It this sense this law is not a pure inertial law. In fact, the contribution of the strongly-dissipative events is non-negligible also in the core of the PDFs of $\Delta u_1$ (but not dominating as in their tails) as can be seen from Fig. \ref{fig:hists}.

Among the main challenges for future work is an experiment similar to that described in references$^{\ref{ref:gkk},\ref{ref:kty}}$ but with sub-Kolmogorov resolution. This includes also the issue of passive scalar. So far, we have pretty crude qualitative results (due to poor resolution and quality of the data) concerning the passive scalar$^{\ref{ref:gkk3}}$, which show the same trends as described above and which raise similar questions concerning the anomalous scaling of passive scalars.

%\bigskip
\newpage
\begin{footnotesize}
\newcounter{refcount}
\begin{list}{$^{\arabic{refcount}}$}{\usecounter{refcount}}

\item T. Gotoh, and R. H. Kraichnan, ``Turbulence and
Tsallis statistics,'' Physica, \textbf{D 193}, 231--244 (2004).\label{ref:gokr}

\item A. N. Kolmogorov,  ``A refinement of previous
hypotheses concerning the local structure of turbulence in a viscous
incompressible fluid at high Reynolds number,'' J. Fluid Mech., 
\textbf{13}, 82--85 (1962).\label{ref:kolm}

\item A. M. Oboukhov, ``Some specific features of
atmospheric turbulence,'' J. Fluid Mech., \textbf{13}, 77--81 (1962).\label{ref:obou}

\item U. Frisch,  \emph{Turbulence --- The Legacy of A. N.
Kolmogorov}, (Cambridge University Press, 1995).\label{ref:frish}

\item A. Tsinober, \emph{An informal conceptual introduction to turbulence}, (Springer, 2009), in press.\label{ref:atsi}

\item J. Schumacher, K. R. Sreenivasan, and V. Yakhot, 
``Asymptotic exponents from low-Reynolds-number flows,'' New J.
Physics, \textbf{9}, 89 (1--19) (2007).\label{ref:ssya}

\item T. S. Lundgren,  ``Turbulent scaling,'' Phys. Fluids, \textbf{20}, 031301/1--10 (2008).\label{ref:lund}

\item V. S. Lvov, and I. Procaccia,  ``Intermittency in
hydrodynamic turbulence as intermediate asymptotics to Kolmogorov scaling,'' 
Phys. Rev. Lett., \textbf{74}, 2690--2693 (1995).\label{ref:lvpr}

\item T. Nakano, T. Gotoh,  and D. Fukayama, ``Roles of
convection, pressure, and dissipation in three-dimensional turbulence,'' 
Phys. Rev., \textbf{E67}, 026316/1--14 (2003).\label{ref:ngfu}

\item J. Qian,  ``Normal and anomalous scaling of
turbulence,'' Phys. Rev., \textbf{E58}, 7325--7329 (1998).\label{ref:qian}

\item D. Ruelle,  ``Conceptual problems of weak and
strong turbulence,'' Phys. Reports, \textbf{103}, 81--85 (1984).\label{ref:ruel}

\item G. Gulitski, M. Kholmyansky, W. Kinzelbach,
B. L\"{u}thi, A. Tsinober, and S. Yorish,  ``Velocity and
temperature derivatives in high-Reynolds-number turbulent flows in
the atmospheric surface layer. Part 1. Facilities, methods and
some general results,'' J. Fluid Mech. {\bf 589}, 57--81 (2007).\label{ref:gkk}

\item M. Kholmyansky, A. Tsinober, and S. Yorish,
 ``Velocity derivatives in the atmospheric surface layer at
$Re_{\lambda }=10^{4}$,'' Phys. Fluids {\bf 13}, 311--314 (2001).\label{ref:kty}

\item M. Kholmyansky, and A. Tsinober, ``Kolmogorov
4/5 law, nonlocality, and sweeping decorrelation hypothesis,'' Phys.
Fluids, \textbf{20}, 041704/1--4 (2008).\label{ref:khts}

\item G. Gulitski, M. Kholmyansky, W. Kinzelbach,
B. L\"{u}thi, A. Tsinober, and S. Yorish,  ``Velocity and
temperature derivatives in high-Reynolds-number turbulent flows in
the atmospheric surface layer. Part 3. Temperature and joint statistics of temperature and velocity derivatives,'' J. Fluid Mech. {\bf 589}, 103--123 (2007).\label{ref:gkk3}

\end{list}
\end{footnotesize}
\end{document}